\documentclass[sigconf,natbib,anonymous=false,review=false]{acmart}
\usepackage[skip=0pt]{caption}
\usepackage{color}
\usepackage{multirow}
\usepackage{dsfont}
\usepackage{relsize}
\usepackage{colortbl}
\usepackage{algorithm,algorithmic}
\usepackage{xcolor}
\usepackage{mathtools}
\usepackage{afterpage}

\usepackage{comment}
\usepackage{amsmath}
\usepackage{accents}
\usepackage{hyperref}
\usepackage{enumerate}
\usepackage[shortlabels,inline]{enumitem}
\usepackage{tikz}

\AtBeginDocument{
 \providecommand\BibTeX{{%
  \normalfont B\kern-0.5em{\scshape i\kern-0.25em b}\kern-0.8em\TeX}}}
\copyrightyear{}
\acmYear{2023}
\setcopyright{None}
\acmConference[Version 1.0]{}{May 3rd, 2023}{}
\acmBooktitle{}
\acmDOI{}
\acmISBN{}
\author{Arian Askari}
\email{a.askari@liacs.leidenuniv.nl}
\affiliation{%
  \institution{Leiden University}
  \country{The Netherlands}
}
\author{Mohammad Aliannejadi}
\email{m.aliannejadi,e.kanoulas}
\affiliation{%
  \institution{University of Amsterdam}
  \country{The Netherlands}
}
\author{Evangelos Kanoulas}
\email{e.kanoulas@uva.nl}
\affiliation{%
  \institution{University of Amsterdam}
  \country{The Netherlands}
}
\author{Suzan Verberne}
\email{s.verberne@liacs.leidenuniv.nl}
\affiliation{%
  \institution{Leiden University}
  \country{The Netherlands}
}
\begin{document}
\title[Generating Synthetic Documents for Cross-Encoder Re-Rankers]{Generating Synthetic Documents for Cross-Encoder Re-Rankers: A Comparative Study of ChatGPT and Human Experts}
\begin{abstract}
We investigate the usefulness of generative Large Language Models (LLMs) in generating training data for cross-encoder re-rankers in a novel direction: generating synthetic documents instead of synthetic queries. We introduce a new dataset, ChatGPT-RetrievalQA, and compare the effectiveness of models fine-tuned on LLM\-generated and human-generated data. Data generated with generative LLMs can be used to augment training data, especially in domains with smaller amounts of labeled data. We build ChatGPT-RetrievalQA based on an existing dataset, human ChatGPT Comparison Corpus (HC3), consisting of public question collections with human responses and answers from ChatGPT. We fine-tune a range of cross-encoder re-rankers on either human-generated or ChatGPT-generated data. Our evaluation on MS MARCO DEV, TREC DL'19, and TREC DL'20 demonstrates that cross-encoder re-ranking models trained on ChatGPT responses are statistically significantly more effective zero-shot re-rankers than those trained on human responses. In a supervised setting, the human-trained re-rankers outperform the LLM-trained re-rankers. Our novel findings suggest that generative LLMs have high potential in generating training data for neural retrieval models. Further work is needed to determine the effect of factually wrong information in the generated responses and test our findings' generalizability with open-source LLMs. We release our data, code, and cross-encoders checkpoints for future work. \footnote{\href{https://github.com/arian-askari/ChatGPT-RetrievalQA}{https://github.com/arian-askari/ChatGPT-RetrievalQA}}
\end{abstract}
\maketitle

\section{Introduction}
Generative large language models (LLMs) such as GPT-3 \cite{brown2020language} and GPT-3.5 (including ChatGPT) have shown remarkable performance in generating realistic text outputs for a variety of tasks such as summarization \cite{zhang2023extractive},
%multimodal reasoning \cite{yang2023mm}, 
machine translation \cite{peng2023towards}, sentiment analysis \cite{susnjak2023applying,wang2023chatgpt}, retrieval interpretability \cite{llordes2023explain}, and stance detection \cite{zhang2022would}. 
Although ChatGPT can produce impressive answers, it is not immune to errors or hallucinations \cite{goldstein2023generative}.
Furthermore, the lack of transparency in the source of information generated by ChatGPT can be a bigger concern in domains such as law, medicine, and science, where accountability and trustworthiness are critical \cite{cascella2023evaluating,sun2023short,sallam2023chatgpt}.
\par
Retrieval models, as opposed to generative models, retrieve the actual (true) information from sources and search engines provide the source of each retrieved item \cite{sanderson2012history}. This is why information retrieval (IR) --- even when generative LLMs are available --- remains an important application, especially in situations where reliability is vital. 
One potential purpose of generative LLMs in IR is to generate training data for retrieval models. Data generated with generative LLMs can be used to augment training data, especially in domains with smaller amounts of labeled data. 
In this paper, we present the ChatGPT-RetrievalQA dataset to address two research questions:
\par
\noindent \textbf{RQ1:} How does the effectiveness of cross-encoder re-rankers fine-tuned on ChatGPT-generated responses compare to those fine-tuned on human-generated responses in both supervised and zero-shot settings?
\par 
\noindent \textbf{RQ2:} How does the effectiveness of using ChatGPT for generating relevant documents differ between specific and general domains?
\par
\begin{figure}
\centering
\scalebox{0.49}{
    \includegraphics[]{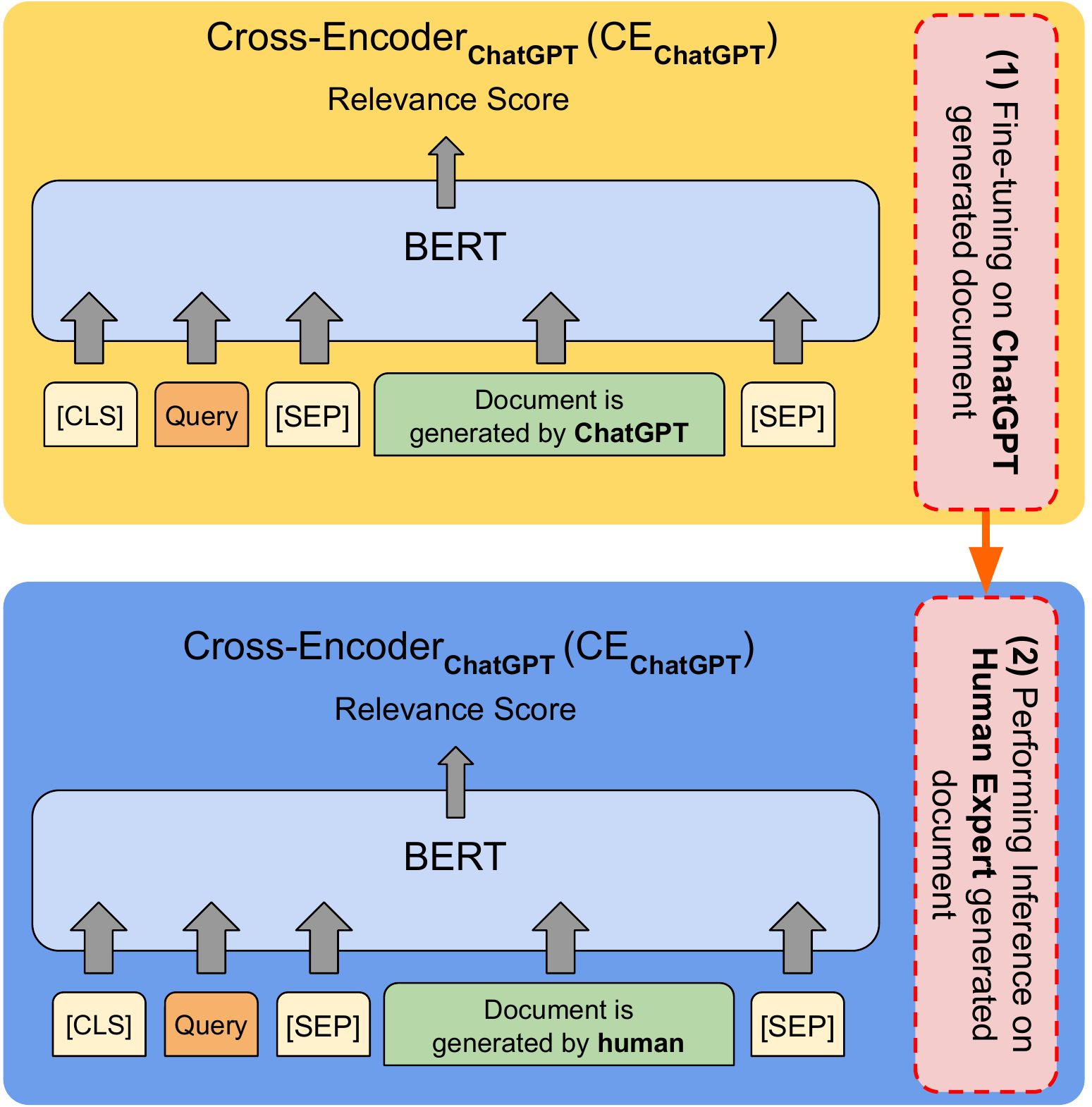}
}
\caption{Our main experimental setup involves above steps.}
\label{fig:setup}
\end{figure}

By answering these questions, we aim to shed light on the potential of using LLMs for data augmentation in cross-encoder re-rankers and the domain dependency of their effectiveness. 
As shown in Figure \ref{fig:setup}, our primary experimental setup involves using CE\textsubscript{ChatGPT}\footnote{We refer to the cross-encoders fine-tuned on ChatGPT-generated and human-generated responses as CE\textsubscript{ChatGPT} and CE\textsubscript{human}, respectively.} for inference (i.e., re-ranking task) on human-generated responses.
\par
Our dataset and analysis provide insights into the benefits and limitations of using generative LLMs for augmenting training data for retrieval models.
\par
Our main contributions in this work are three-fold:
\begin{enumerate*}[label=(\roman*)]
    \item We release the ChatGPT-RetrievalQA dataset, which is based on the HC3 dataset \cite{guo2023close} but is designed specifically for information retrieval tasks in both full-ranking and re-ranking setups. This dataset contains $24,322$ queries, $26,882$ responses generated by ChatGPT, and $58,546$ human-generated responses.
    \item We fine-tune cross-encoder re-rankers on both the human- and ChatGPT-generated responses, evaluating their performance on our dataset in a supervised setting. We also show the effectiveness of the ChatGPT-trained models in a zero-shot evaluation on the MS MARCO-passage collection and the TREC Deep Learning tracks.
    \item We conduct an analysis of the effectiveness of ChatGPT-trained cross-encoders on different domains and show that human-trained models are slightly more effective in domain-specific tasks, e.g., in the medicine domain.
\end{enumerate*}
Our novel findings highlight the potential of using generative LLMs like ChatGPT for generating high-quality responses in information retrieval tasks in order to create training datasets.
\section{Related Work}
InPars \cite{bonifacio2022inpars}, Promptagator \cite{dai2022promptagator}, and InPars-v2 \cite{jeronymo2023inpars} have utilized LLMs to generate synthetic queries given documents. Particularly, InPars-v2 \cite{jeronymo2023inpars} achieves state-of-the-art results on the BEIR dataset in a zero-shot setting by using an open-source language model, GPT-J-6B \cite{gpt-j} and a powerful external re-ranker, MonoT5-MSMARCO \cite{nogueira2020document} to filter the top-10k high-quality pairs of synthetic query-document pairs for data augmentation. In contrast, we use \textit{documents} (passages) generated by ChatGPT given a query -- the reverse from InPars-v2. Document generation for given queries as a source for data augmentation has not been explored in prior work. 
\par
% doc2query vs. query2doc
We believe that exploring this reverse direction is important as it allows us to augment training data with a focus on user behavior and query logs rather than the (static) document collection itself. This can improve the effectiveness of re-rankers by augmenting the training data with synthetic documents according to the queries that actual search engine users are searching for,
increasing the diversity of the training data, while allowing the rankers to better generalize to new queries.
\par
\citet{guo2023close} use public question-answering datasets (see below) and prompt the questions to the ChatGPT user interface for generating answers. The goal of the HC3 dataset is to linguistically compare human and ChatGPT responses and explore the possibility of distinguishing between responses generated by ChatGPT and those written by humans. The HC3 dataset contains questions (i.e., queries) from four different domains: medicine (Medical Dialog \cite{chen2020meddialog}), finance (FiQA \cite{maia201818}), Wikipedia (WikiQA \cite{yang2015wikiqa} and Wiki\_csai \cite{guo2023close}), and Reddit (ELI5 \cite{fan2019eli5}). 
While there is no study on generating documents to augment training data, a more recent study, QuerytoDoc \cite{wang2023query2doc}, generates documents for query expansion, which is out of the scope of augmenting data for information retrieval.
Furthermore, there are various recent studies on ChatGPT with a focus on ranking and retrieval but to the best of our knowledge, none of them focus on data augmentation by generating relevant documents. Examples of recent studies are the one by Faggioli et al \cite{faggioli2023perspectives}, who study if ChatGPT can be used for generating relevance labels, and \citet{sun2023chatgpt}, who assess whether ChatGPT is good at searching by giving it a query and set of candidate documents and asking for re-ranking.
\begin{table}[t]
\caption{Statistic on the size of Train, Validation, and Test sets across domains for evaluation of cross-Encoders.}
\label{table:stat}
% \scalebox{0.85}{
    \begin{tabular}{lccc}
            \toprule
            \multirow{2}{*}{Domain}  & \multicolumn{3}{c}{\# of queries}   \\ 
                                  & Train set      & Validation set & Test set  \\ \midrule
            All                            & 16788      & 606        & 6928  \\ 
            Medicine: Meddialog \cite{chen2020meddialog}                      & 862        & 31         & 355   \\ 
            Finance: FiQA \cite{maia201818}                    & 2715       & 98         & 1120  \\ 
            Reddit: ELI5 \cite{fan2019eli5}                        & 11809      & 427        & 4876  \\
            Wikipedia: openQA \cite{yang2015wikiqa}    & 820        & 29         & 338   \\ 
            Wikipedia: csai \cite{guo2023close} & 582        & 21         & 239  \\
          \bottomrule
    \end{tabular}
% }
\end{table}
\section{Dataset}
Our ChatGPT-RetrievalQA dataset is based on the HC3 dataset produced by \citet{guo2023close}, which contains $24,322$ queries and $26,882$ ChatGPT-generated responses, as well as $58,546$ human-generated responses. There is on average one ChatGPT-generated and $2.4$ human-generated response per query. To build the ChatGPT-RetrievalQA dataset for retrieval and ranking experiments (an experimental setup different from \cite{guo2023close}), we convert the dataset files to a format similar to the MSMarco dataset \cite{nguyen2016ms}, in both full-ranking and re-ranking setups.\footnote{This allows  for easy reuse of available scripts on MS MARCO.} We divide the data into training, validation, and test sets.
\begin{table*}[ht]
    \centering
    \caption{Comparing the effectiveness of cross-encoder re-rankers fine-tuned on human and ChatGPT responses in supervised and zero-shot settings. $\dagger$ indicates that a CE achieves statistically significant improvement for that dataset among all of the cross-encoder re-rankers and BM25 on that dataset. Statistical significance was measured with a paired t-test ($p<0.05$) with Bonferroni correction for multiple testing. The cutoff for MAP, NDCG, and MRR are 1000, 10, and 10.}
    \label{table:results}
    \begin{tabular}{lccc|ccc|ccc|ccc}
        \toprule
        & \multicolumn{9}{c|}{Zero-shot setting} & \multicolumn{3}{c}{Supervised setting} \\ \midrule
        \multirow{2}{*}{Model} & \multicolumn{3}{c|}{TREC DL'19} & \multicolumn{3}{c|}{TREC DL'20} & \multicolumn{3}{c|}{MS MARCO DEV}  & \multicolumn{3}{c}{ChatGPT-RetrievalQA \textbf{(Ours)}} \\
        & MAP & NDCG & MRR & MAP & NDCG & MRR  & MAP & NDCG & MRR  & MAP & NDCG & MRR  \\
        \midrule
        BM25 & \textbf{.377} &  .506 & .858 & .286 &  .480 &  .819 & .195 &  .234 &  .187 & .143 &  .184 &  .240 \\
        \midrule
        MiniLM\textsubscript{human} & .326 & .451 & .833 & .269 & .376 & .913 & .130 & .155 & .118 &  \textbf{.310}$\dagger$ & \textbf{.384}$\dagger$ & \textbf{.460}$\dagger$  \\
        MiniLM\textsubscript{ChatGPT} & .342$\dagger$ & \textbf{.510}$\dagger$ & .903 & \textbf{.344}$\dagger$ & \textbf{.539}$\dagger$ & \textbf{.978}$\dagger$ & \textbf{.226}$\dagger$ & \textbf{.267}$\dagger$ & \textbf{.218}$\dagger$ & .294 & .362 & .444 \\
        TinyBERT\textsubscript{human} & .294 & .360 & .741 & .277 & .364 & .791 & .128 & .154 & .116 & .244 & .310 & .367 \\
        TinyBERT\textsubscript{ChatGPT} & .328 & .488 & \textbf{.942}$\dagger$ & .303 & .460 & .972 & .194 & .231 & .185 & .231 & .291 & .358\\
        \bottomrule
    \end{tabular}
\end{table*}
\par
To facilitate training, we provide training triples files in TSV format, including both textual and ID-based representations, where the structure of each line is composed of `query, positive passage, negative passage' and `qid, positive pid, negative pid'. We consider the actual response by ChatGPT or human as the relevant answer and we randomly sample $1000$ negative answers for each query similar to MS MARCO. In addition, we provide the top 1000 documents, ranked by BM25, per query to enable re-ranking studies. Table \ref{table:stat} shows the size of the train, validation, and test sets for each domain.
\section{Methods}
\subsection{First-stage ranker: BM25}
Lexical retrievers use word overlap to produce the relevance score between a document and a query. Several lexical approaches have been developed in the past, such as vector space models, Okapi BM25 \cite{robertson1994some}, and query likelihood. We use BM25 as the first-stage ranker because of its popularity and effectiveness. BM25 calculates a score for a query--document pair based on the statistics of the words that overlap between them:
\begin{equation}
        \resizebox{0.88\hsize}{!}{$s_{lex}(q,d) = BM25(q,d) = 
        \sum_{t \in q \cap d }{rsj_t . \frac{tf_{t,d}}{tf_{t,d} + k_{1} \{ (1-b) + b \frac{|d|}{l} \} }}$}
\end{equation}
where $t$ is a term, $tf_{t,d}$ is the frequency of $t$ in document $d$, $rsj_t$ is the Robertson-Spärck Jones weight \cite{robertson1994some} of $t$, and $l$ is the average document length. $k_1$ and $b$ are parameters.
\subsection{Cross-encoder re-rankers}
The common approach to employ pre-trained Transformer models with a cross-encoder architecture in a re-ranking setup is by concatenating the input sequences of query and passage. This method, known as MonoBERT or CE\textsubscript{CAT}, is illustrated in Figure \ref{fig:setup} and has been utilized in several studies. In CE\textsubscript{CAT}, the sequences of query words $q_1:q_m$ and passage words $p_1:p_n$ are joined with the [SEP] token, and the ranking model of CE\textsubscript{CAT} calculates the score for the representation of the [CLS] token obtained by cross-encoder (CE) using a single linear layer W\textsubscript{s}:
\begin{equation}
    \resizebox{0.88\hsize}{!}{$CE_{CAT}(q_{1:m},p_{1:n}) = CE([CLS]\,q\,[SEP]\,p\,[SEP]) * W_s$}
\end{equation}
We use CE\textsubscript{CAT} as our cross-encoder re-ranker with a re-ranking depth of $1000$. In our experiments, both CE\textsubscript{ChatGPT} and CE\textsubscript{human} follow the above design.
\section{Experimental design}
\textbf{Evaluation setup.} We conduct our zero-shot evaluation experiments on the MS MARCO-passage collection \cite{nguyen2016ms} and the two TREC Deep Learning tracks (TREC-DL'19 and DL'20) \cite{craswell2020overview,craswell2021overview}. 
The evaluation metrics used are MAP@1000, NDCG@10, and MRR@10, which are standard for these datasets, to make our results comparable to previously published and upcoming research \cite{craswell2021overview,craswell2020overview}. The MS MARCO-passage dataset contains about $8.8$ million passages (average length: $73.1$ words) and about $1$ million natural language queries (average length: $7.5$ words) and has been extensively used to train deep language models for ranking. Following prior work on MS MARCO \cite{khattab2020colbert,lin2021pretrained,macavaney2020expansion,zhuang2021tilde,zhuang2021deep}, we only use the dev set ($\sim 7k$ queries) for our empirical evaluation. The TREC DL'19 and DL'20 collections share the passage corpus of MS MARCO and have $43$ and $54$ queries respectively with much more relevant documents per query. We measure the same metrics in the supervised setting on the test set of ChatGPT-RetrievalQA. The average length of human responses is $142.5$ words, and $198.1$ words for ChatGPT in the ChatGPT-RetrievalQA dataset.
\par
\textbf{Training configuration.}
We use the Huggingface library \cite{wolf2019huggingface}, and PyTorch \cite{paszke2017automatic} for the cross-encoder re-ranking training and inference.
Following prior work \cite{hofstatter2020improving}, we use the Adam \cite{kingma2014adam} optimizer with a learning rate of $7*10^{-6}$ for all cross-encoder layers, regardless of the number of layers trained. We use a training batch size of 32. For all cross-encoder re-rankers, we use the cross-entropy loss \cite{zhang2018generalized}. For the lexical retrieval with BM25, we use the similarity function of Elasticsearch \cite{elasticsearch2023elasticsearch}. We cap the query length at $30$ tokens and the passage length at $200$ tokens following prior work \cite{hofstatter2020improving,AskariECIR2023}.
\begin{table}[]
\centering
\caption{Comparing the effectiveness of CE\textsubscript{C} and CE\textsubscript{H} in supervised setting across different domains where CE, C, and H refer to the MiniLM, human, and ChatGPT. The OpenQA and Wiki\_csai datasets are in Wikipedia domain.}
\label{table:domain_results}
    \begin{tabular}{ccccc}
    \toprule
    Domain & Model & MAP@1K & NDCG@10 & MRR@10\\ 
    \midrule
    \multirow{2}{*}{All}        & CE\textsubscript{H}     & \textbf{.310 }    & \textbf{.384}    & \textbf{.460}     \\
    & CE\textsubscript{C}    & .294     & .362    & .444     \\ \midrule
    \multirow{2}{*}{Medicine \cite{chen2020meddialog}}                  & CE\textsubscript{H}      & \textbf{.397}     & \textbf{.419}    & \textbf{.395 }     \\
    &  CE\textsubscript{C}    & .379     & .400    & .377      \\ \midrule
    \multirow{2}{*}{Finance \cite{maia201818}}                  & CE\textsubscript{H}      & \textbf{.257}     & \textbf{.399}    & \textbf{.251}      \\ 
     & CE\textsubscript{C}    & .250     & .368    & .245      \\ \midrule
    \multirow{2}{*}{Reddit \cite{fan2019eli5}}                & CE\textsubscript{H}      & \textbf{.323}     & \textbf{.418}    & \textbf{.543}      \\ 
    & CE\textsubscript{C}    & .302     & .391    & .522      \\ \midrule
    \multirow{2}{*}{OpenQA \cite{yang2015wikiqa} }                & CE\textsubscript{H}      & .322     & \textbf{.345}    & .320      \\
     & CE\textsubscript{C}    & \textbf{.331}     & .341    & \textbf{.328}      \\ \midrule
    \multirow{2}{*}{Wiki\_csai \cite{guo2023close}}        & CE\textsubscript{H}      & .149     & .152    & .135      \\
     & CE\textsubscript{C}    & \textbf{.163}     & \textbf{.159}    & \textbf{.144 } \\
    \bottomrule
    \end{tabular}
\end{table}
\section{Results}
\subsection{Main results (RQ1)}
Table \ref{table:results} shows a comparison of the effectiveness of CE\textsubscript{human} and CE\textsubscript{ChatGPT}. Please note that for both models, during inference, we evaluate their effectiveness in retrieving \textit{human} responses in the supervised or zero-shot settings.
We choose MiniLM (12 layers version) \cite{wang2020minilm} for the experiments due to its competitive results in comparison to BERT re-ranker \cite{AskariECIR2023} while being three times smaller and six times faster. In addition, we conduct experiments with TinyBERT (2 layers version) \cite{jiao-etal-2020-tinybert} to assess the generalizability of our results.
\par
In the \textbf{supervised setting} where we evaluate the on test set queries with human documents of our ChatGPT-RetrievalQA dataset, MiniLM\textsubscript{human} significantly outperforms all of the other cross-encoders re-rankers. Although lower than the human-trained models, MiniLM\textsubscript{ChatGPT} and TinyBERT\textsubscript{ChatGPT} still outperform the strong baseline \cite{bonifacio2022inpars}, BM25 \cite{robertson1994some}, statistically significantly by a large margin in this setting.
\par
In the \textbf{zero-shot setting}, the MiniLM\textsubscript{ChatGPT} consistently outperforms the other cross-encoder re-rankers including MiniLM\textsubscript{human} and BM25 significantly across the TREC DL'20 and MS MARCO DEV. However, on TREC DL'19, BM25 achieves the highest effectiveness for MAP@1000, MiniLM\textsubscript{ChatGPT} for NDCG@10, and TinyBERT\textsubscript{ChatGPT} for MRR@10. Overall, we can see the models fine-tuned on ChatGPT-generated responses are significantly more effective in the zero-shot setting compared to those fine-tuned on human-generated responses.
\subsection{Domain-level re-ranker effectiveness (RQ2)}
Table \ref{table:domain_results} shows the effectiveness of MiniLM\textsubscript{human} and MiniLM\textsubscript{ChatGPT} re-rankers in the supervised settings -- on the test set of our dataset -- across all of the domains including Medicine, Finance, Reddit, and Wikipedia.
Overall, the results show that MiniLM\textsubscript{human} achieves higher effectiveness than MiniLM\textsubscript{ChatGPT}
% with higher MAP@1k, NDCG@10, and MRR@10 scores
for all domains except Wikipedia. However, the difference in effectiveness is small, and MiniLM\textsubscript{ChatGPT} still achieves a reasonable level of effectiveness. In the Finance domain, both MiniLM\textsubscript{human} and MiniLM\textsubscript{ChatGPT} achieve relatively low effectiveness compared to other domains. In the Wikipedia domain, MiniLM\textsubscript{human} and MiniLM\textsubscript{ChatGPT} achieve relatively similar levels of effectiveness. In the Medicine domain, the CE\textsubscript{human} shows the highest effectiveness. Overall, these results suggest that MiniLM\textsubscript{human} performs more effective in supervised settings, particularly in specific domains such as Medicine, even though the difference in performance is small.
\section{Discussion}
\textbf{Data overlap.} It is worth noting that in the supervised setting, the collection of documents used for training and testing is shared for CE\textsubscript{human} re-rankers. Therefore, some documents may be seen during both training and evaluation. This setup is very common when working with human-assessed data, and similar to the MS MARCO dataset \cite{nguyen2016ms}. The shared collection could be a potential benefit for CE\textsubscript{human} re-rankers in the supervised setting, as the models may have already seen some of the documents during training. To further investigate this hypothesis, it would be worth exploring a different setup in future work in which the collection of documents is completely separated between the training and test sets. %We leave this to future work.
\par
\textbf{Effectiveness of BM25.} 
Table \ref{table:bm25} shows an analysis of the effectiveness of BM25 on human and ChatGPT-generated responses in the train, and test sets. BM25 is less effective for human-generated responses than for ChatGPT-generated responses on the train and test
% , and validation
sets, as evidenced by lower scores for all metrics. 
We observed the same pattern for the validation set.
These results suggest that the task of retrieving human-generated responses is more challenging for BM25 than for ChatGPT-generated responses.
This is probably related to the lexical overlap discussed below.
\par
\textbf{Queries without label.} 
In Table \ref{table:seen_queries}, we investigate a common scenario in real-world search engines where query logs and a collection of human-generated documents are available, and there are not any judged documents for part or all of the query logs. To simulate and analyze this situation, we evaluate CE\textsubscript{ChatGPT} on the seen queries of the train set and unseen documents of the human-generated documents collection.
% Our results, as shown in Table \ref{table:seen_queries}, demonstrate that 
Table \ref{table:seen_queries} shows that
CE\textsubscript{ChatGPT} rankers are fairly effective in this scenario. Especially, they are more effective than BM25 in the same setup, in that the NDCG@10 for MiniLM\textsubscript{ChatGPT} is 0.388 and 0.202 for BM25 (see the third row of Table \ref{table:bm25}). This suggests the potential of augmenting training data with generative LLMs for fine-tuning models to effectively re-rank sourced and reliable human-generated documents from the corpus given the query logs where there are no judged documents for the queries.
\begin{table}[t]
\centering
\caption{Analyzing the effectiveness of BM25 on human/ChatGPT responses in train, validation, and test set.}
\label{table:bm25}
    \begin{tabular}{lcccc}
    \toprule
    Split & Source & MAP@1K & NDCG@10 & Recall@1K   \\ \midrule
    \multirow{2}{*}{Test}            & human                & .143     & .184    & .520        \\
                    & ChatGPT              & .370     & .396    & .898        \\
                    \midrule 
    \multirow{2}{*}{Train}           & human                & .158     & .202    & .560      \\
                    & ChatGPT              & .413     & .443    & .903      \\
    \bottomrule
    \end{tabular}
\end{table}
\par
\textbf{Lexical overlap.}
Our data analysis reveals that ChatGPT-generated responses have a slightly higher lexical overlap than human-generated responses with the queries. 
The average percentage of query words that occur in ChatGPT-generated responses is 34.6\%, compared to 25.5\% for humans. The Q1, median, and Q3 are also on average 7\% points higher for ChatGPT compared to human responses. We suspect that this higher lexical overlap compared to the human response happens because ChatGPT often repeats the question or query in the response, and it tends to generate lengthier responses compared to human which increase the chance of having mutual words with the query.
It is noteworthy to mention that lexical overlap is not the best indicator of response quality for fine-tuning effective cross-encoders, as there may be cases where responses with low lexical overlap are still relevant and informative, especially in question-answering tasks.
\begin{table}[]
\centering
\caption{Analyzing the effectiveness of CEs\textsubscript{ChatGPT} on the seen queries of the train set and unseen documents of human-generated documents collection.}
\label{table:seen_queries}
    \begin{tabular}{lccc}
    \toprule
    Model & MAP@1K & NDCG@10 & MRR@10   \\ \midrule
    MiniLM\textsubscript{ChatGPT}                & .318     & .388    & .510        \\
    TinyBERT\textsubscript{ChatGPT}              & .254     & .318    & .420        \\
    \bottomrule
    \end{tabular}
\end{table}
\section{Conclusion}
In this paper, we have presented an analysis of the effectiveness of fine-tuning cross-encoders on human-generated responses versus ChatGPT-generated responses. 
Our results show that the cross-encoder\textsubscript{ChatGPT} is more effective than cross-encoder\textsubscript{human} in the zero-shot setting while MiniLM human is slightly more effective in the supervised setting and this is consistent across different domains. Furthermore, we show that BM25 is less effective on human-generated responses than on ChatGPT-generated responses, indicating that human-generated responses are more challenging to match with queries than ChatGPT-generated responses. 
Overall, our findings suggest that ChatGPT-generated responses are more useful than human-generated responses for training effective zero-shot re-ranker, at least based on our dataset and experiments, and highlight the potential of using generative LLMs for generating effective and useful responses for creating training datasets in natural language processing tasks.
Our study can be particularly advantageous for domain-specific tasks where relying on LLM-generated output as a direct response to a user query can be risky. Our results confirm that it is possible to train effective cross-encoder re-rankers by training them on ChatGPT-generated responses even for domain-specific queries.
Further work is needed to determine the effect of factually wrong information in the generated responses and to test the generalizability of our findings on open-source LLMs.
\bibliographystyle{ACM-Reference-Format}
\balance
\bibliography{references}


\begin{thebibliography}{43}

%%% the Web address.

\ifx \showCODEN    \undefined \def \showCODEN     #1{\unskip}     \fi
\ifx \showDOI      \undefined \def \showDOI       #1{#1}\fi
\ifx \showISBNx    \undefined \def \showISBNx     #1{\unskip}     \fi
\ifx \showISBNxiii \undefined \def \showISBNxiii  #1{\unskip}     \fi
\ifx \showISSN     \undefined \def \showISSN      #1{\unskip}     \fi
\ifx \showLCCN     \undefined \def \showLCCN      #1{\unskip}     \fi
\ifx \shownote     \undefined \def \shownote      #1{#1}          \fi
\ifx \showarticletitle \undefined \def \showarticletitle #1{#1}   \fi
\ifx \showURL      \undefined \def \showURL       {\relax}        \fi
% invisible to TeX
\providecommand\bibfield[2]{#2}
\providecommand\bibinfo[2]{#2}
\providecommand\natexlab[1]{#1}
\providecommand\showeprint[2][]{arXiv:#2}

\bibitem[\protect\citeauthoryear{Askari, Abolghasemi, Pasi, Kraaij, and
  Verberne}{Askari et~al\mbox{.}}{2023}]%
        {AskariECIR2023}
\bibfield{author}{\bibinfo{person}{Arian Askari}, \bibinfo{person}{Amin
  Abolghasemi}, \bibinfo{person}{Gabriella Pasi}, \bibinfo{person}{Wessel
  Kraaij}, {and} \bibinfo{person}{Suzan Verberne}.}
  \bibinfo{year}{2023}\natexlab{}.
\newblock \showarticletitle{Injecting the BM25 Score as Text Improves
  BERT-Based Re-rankers}. In \bibinfo{booktitle}{\emph{Advances in Information
  Retrieval}}. \bibinfo{publisher}{Springer Nature Switzerland},
  \bibinfo{address}{Cham}, \bibinfo{pages}{66--83}.
\newblock
\showISBNx{978-3-031-28244-7}


\bibitem[\protect\citeauthoryear{Bonifacio, Abonizio, Fadaee, and
  Nogueira}{Bonifacio et~al\mbox{.}}{2022}]%
        {bonifacio2022inpars}
\bibfield{author}{\bibinfo{person}{Luiz Bonifacio}, \bibinfo{person}{Hugo
  Abonizio}, \bibinfo{person}{Marzieh Fadaee}, {and} \bibinfo{person}{Rodrigo
  Nogueira}.} \bibinfo{year}{2022}\natexlab{}.
\newblock \showarticletitle{Inpars: Data augmentation for information retrieval
  using large language models}.
\newblock \bibinfo{journal}{\emph{arXiv preprint arXiv:2202.05144}}
  (\bibinfo{year}{2022}).
\newblock


\bibitem[\protect\citeauthoryear{Brown, Mann, Ryder, Subbiah, Kaplan, Dhariwal,
  Neelakantan, Shyam, Sastry, Askell, et~al\mbox{.}}{Brown
  et~al\mbox{.}}{2020}]%
        {brown2020language}
\bibfield{author}{\bibinfo{person}{Tom Brown}, \bibinfo{person}{Benjamin Mann},
  \bibinfo{person}{Nick Ryder}, \bibinfo{person}{Melanie Subbiah},
  \bibinfo{person}{Jared~D Kaplan}, \bibinfo{person}{Prafulla Dhariwal},
  \bibinfo{person}{Arvind Neelakantan}, \bibinfo{person}{Pranav Shyam},
  \bibinfo{person}{Girish Sastry}, \bibinfo{person}{Amanda Askell},
  {et~al\mbox{.}}} \bibinfo{year}{2020}\natexlab{}.
\newblock \showarticletitle{Language models are few-shot learners}.
\newblock \bibinfo{journal}{\emph{Advances in neural information processing
  systems}}  \bibinfo{volume}{33} (\bibinfo{year}{2020}),
  \bibinfo{pages}{1877--1901}.
\newblock


\bibitem[\protect\citeauthoryear{Cascella, Montomoli, Bellini, and
  Bignami}{Cascella et~al\mbox{.}}{2023}]%
        {cascella2023evaluating}
\bibfield{author}{\bibinfo{person}{Marco Cascella}, \bibinfo{person}{Jonathan
  Montomoli}, \bibinfo{person}{Valentina Bellini}, {and} \bibinfo{person}{Elena
  Bignami}.} \bibinfo{year}{2023}\natexlab{}.
\newblock \showarticletitle{Evaluating the feasibility of ChatGPT in
  healthcare: an analysis of multiple clinical and research scenarios}.
\newblock \bibinfo{journal}{\emph{Journal of Medical Systems}}
  \bibinfo{volume}{47}, \bibinfo{number}{1} (\bibinfo{year}{2023}),
  \bibinfo{pages}{1--5}.
\newblock


\bibitem[\protect\citeauthoryear{Chen, Ju, Dong, Fang, Wang, Yang, Zeng, Zhang,
  Zhang, Zhou, et~al\mbox{.}}{Chen et~al\mbox{.}}{2020}]%
        {chen2020meddialog}
\bibfield{author}{\bibinfo{person}{Shu Chen}, \bibinfo{person}{Zeqian Ju},
  \bibinfo{person}{Xiangyu Dong}, \bibinfo{person}{Hongchao Fang},
  \bibinfo{person}{Sicheng Wang}, \bibinfo{person}{Yue Yang},
  \bibinfo{person}{Jiaqi Zeng}, \bibinfo{person}{Ruisi Zhang},
  \bibinfo{person}{Ruoyu Zhang}, \bibinfo{person}{Meng Zhou}, {et~al\mbox{.}}}
  \bibinfo{year}{2020}\natexlab{}.
\newblock \showarticletitle{Meddialog: a large-scale medical dialogue dataset}.
\newblock \bibinfo{journal}{\emph{arXiv preprint arXiv:2004.03329}}
  (\bibinfo{year}{2020}).
\newblock


\bibitem[\protect\citeauthoryear{Craswell, Mitra, Yilmaz, and Campos}{Craswell
  et~al\mbox{.}}{2021}]%
        {craswell2021overview}
\bibfield{author}{\bibinfo{person}{Nick Craswell}, \bibinfo{person}{Bhaskar
  Mitra}, \bibinfo{person}{Emine Yilmaz}, {and} \bibinfo{person}{Daniel
  Campos}.} \bibinfo{year}{2021}\natexlab{}.
\newblock \showarticletitle{Overview of the TREC 2020 deep learning track}.
\newblock \bibinfo{journal}{\emph{arXiv preprint arXiv:2102.07662}}
  (\bibinfo{year}{2021}).
\newblock


\bibitem[\protect\citeauthoryear{Craswell, Mitra, Yilmaz, Campos, and
  Voorhees}{Craswell et~al\mbox{.}}{2020}]%
        {craswell2020overview}
\bibfield{author}{\bibinfo{person}{Nick Craswell}, \bibinfo{person}{Bhaskar
  Mitra}, \bibinfo{person}{Emine Yilmaz}, \bibinfo{person}{Daniel Campos},
  {and} \bibinfo{person}{Ellen~M Voorhees}.} \bibinfo{year}{2020}\natexlab{}.
\newblock \showarticletitle{Overview of the TREC 2019 deep learning track}.
\newblock \bibinfo{journal}{\emph{arXiv preprint arXiv:2003.07820}}
  (\bibinfo{year}{2020}).
\newblock


\bibitem[\protect\citeauthoryear{Dai, Zhao, Ma, Luan, Ni, Lu, Bakalov, Guu,
  Hall, and Chang}{Dai et~al\mbox{.}}{2022}]%
        {dai2022promptagator}
\bibfield{author}{\bibinfo{person}{Zhuyun Dai}, \bibinfo{person}{Vincent~Y
  Zhao}, \bibinfo{person}{Ji Ma}, \bibinfo{person}{Yi Luan},
  \bibinfo{person}{Jianmo Ni}, \bibinfo{person}{Jing Lu},
  \bibinfo{person}{Anton Bakalov}, \bibinfo{person}{Kelvin Guu},
  \bibinfo{person}{Keith~B Hall}, {and} \bibinfo{person}{Ming-Wei Chang}.}
  \bibinfo{year}{2022}\natexlab{}.
\newblock \showarticletitle{Promptagator: Few-shot dense retrieval from 8
  examples}.
\newblock \bibinfo{journal}{\emph{arXiv preprint arXiv:2209.11755}}
  (\bibinfo{year}{2022}).
\newblock


\bibitem[\protect\citeauthoryear{Elasticsearch}{Elasticsearch}{2023}]%
        {elasticsearch2023elasticsearch}
\bibfield{author}{\bibinfo{person}{BV Elasticsearch}.}
  \bibinfo{year}{2023}\natexlab{}.
\newblock \showarticletitle{Elasticsearch}.
\newblock \bibinfo{journal}{\emph{software], version}} \bibinfo{volume}{6},
  \bibinfo{number}{1} (\bibinfo{year}{2023}).
\newblock


\bibitem[\protect\citeauthoryear{Faggioli, Dietz, Clarke, Demartini, Hagen,
  Hauff, Kando, Kanoulas, Potthast, Stein, et~al\mbox{.}}{Faggioli
  et~al\mbox{.}}{2023}]%
        {faggioli2023perspectives}
\bibfield{author}{\bibinfo{person}{Guglielmo Faggioli}, \bibinfo{person}{Laura
  Dietz}, \bibinfo{person}{Charles Clarke}, \bibinfo{person}{Gianluca
  Demartini}, \bibinfo{person}{Matthias Hagen}, \bibinfo{person}{Claudia
  Hauff}, \bibinfo{person}{Noriko Kando}, \bibinfo{person}{Evangelos Kanoulas},
  \bibinfo{person}{Martin Potthast}, \bibinfo{person}{Benno Stein},
  {et~al\mbox{.}}} \bibinfo{year}{2023}\natexlab{}.
\newblock \showarticletitle{Perspectives on Large Language Models for Relevance
  Judgment}.
\newblock \bibinfo{journal}{\emph{arXiv preprint arXiv:2304.09161}}
  (\bibinfo{year}{2023}).
\newblock


\bibitem[\protect\citeauthoryear{Fan, Jernite, Perez, Grangier, Weston, and
  Auli}{Fan et~al\mbox{.}}{2019}]%
        {fan2019eli5}
\bibfield{author}{\bibinfo{person}{Angela Fan}, \bibinfo{person}{Yacine
  Jernite}, \bibinfo{person}{Ethan Perez}, \bibinfo{person}{David Grangier},
  \bibinfo{person}{Jason Weston}, {and} \bibinfo{person}{Michael Auli}.}
  \bibinfo{year}{2019}\natexlab{}.
\newblock \showarticletitle{ELI5: Long form question answering}.
\newblock \bibinfo{journal}{\emph{arXiv preprint arXiv:1907.09190}}
  (\bibinfo{year}{2019}).
\newblock


\bibitem[\protect\citeauthoryear{Goldstein, Sastry, Musser, DiResta, Gentzel,
  and Sedova}{Goldstein et~al\mbox{.}}{2023}]%
        {goldstein2023generative}
\bibfield{author}{\bibinfo{person}{Josh~A Goldstein}, \bibinfo{person}{Girish
  Sastry}, \bibinfo{person}{Micah Musser}, \bibinfo{person}{Renee DiResta},
  \bibinfo{person}{Matthew Gentzel}, {and} \bibinfo{person}{Katerina Sedova}.}
  \bibinfo{year}{2023}\natexlab{}.
\newblock \showarticletitle{Generative Language Models and Automated Influence
  Operations: Emerging Threats and Potential Mitigations}.
\newblock \bibinfo{journal}{\emph{arXiv preprint arXiv:2301.04246}}
  (\bibinfo{year}{2023}).
\newblock


\bibitem[\protect\citeauthoryear{Guo, Zhang, Wang, Jiang, Nie, Ding, Yue, and
  Wu}{Guo et~al\mbox{.}}{2023}]%
        {guo2023close}
\bibfield{author}{\bibinfo{person}{Biyang Guo}, \bibinfo{person}{Xin Zhang},
  \bibinfo{person}{Ziyuan Wang}, \bibinfo{person}{Minqi Jiang},
  \bibinfo{person}{Jinran Nie}, \bibinfo{person}{Yuxuan Ding},
  \bibinfo{person}{Jianwei Yue}, {and} \bibinfo{person}{Yupeng Wu}.}
  \bibinfo{year}{2023}\natexlab{}.
\newblock \showarticletitle{How Close is ChatGPT to Human Experts? Comparison
  Corpus, Evaluation, and Detection}.
\newblock \bibinfo{journal}{\emph{arXiv preprint arXiv:2301.07597}}
  (\bibinfo{year}{2023}).
\newblock


\bibitem[\protect\citeauthoryear{Hofst{\"a}tter, Althammer, Schr{\"o}der,
  Sertkan, and Hanbury}{Hofst{\"a}tter et~al\mbox{.}}{2020}]%
        {hofstatter2020improving}
\bibfield{author}{\bibinfo{person}{Sebastian Hofst{\"a}tter},
  \bibinfo{person}{Sophia Althammer}, \bibinfo{person}{Michael Schr{\"o}der},
  \bibinfo{person}{Mete Sertkan}, {and} \bibinfo{person}{Allan Hanbury}.}
  \bibinfo{year}{2020}\natexlab{}.
\newblock \showarticletitle{Improving efficient neural ranking models with
  cross-architecture knowledge distillation}.
\newblock \bibinfo{journal}{\emph{arXiv preprint arXiv:2010.02666}}
  (\bibinfo{year}{2020}).
\newblock


\bibitem[\protect\citeauthoryear{Jeronymo, Bonifacio, Abonizio, Fadaee, Lotufo,
  Zavrel, and Nogueira}{Jeronymo et~al\mbox{.}}{2023}]%
        {jeronymo2023inpars}
\bibfield{author}{\bibinfo{person}{Vitor Jeronymo}, \bibinfo{person}{Luiz
  Bonifacio}, \bibinfo{person}{Hugo Abonizio}, \bibinfo{person}{Marzieh
  Fadaee}, \bibinfo{person}{Roberto Lotufo}, \bibinfo{person}{Jakub Zavrel},
  {and} \bibinfo{person}{Rodrigo Nogueira}.} \bibinfo{year}{2023}\natexlab{}.
\newblock \showarticletitle{InPars-v2: Large Language Models as Efficient
  Dataset Generators for Information Retrieval}.
\newblock \bibinfo{journal}{\emph{arXiv preprint arXiv:2301.01820}}
  (\bibinfo{year}{2023}).
\newblock


\bibitem[\protect\citeauthoryear{Jiao, Yin, Shang, Jiang, Chen, Li, Wang, and
  Liu}{Jiao et~al\mbox{.}}{2020}]%
        {jiao-etal-2020-tinybert}
\bibfield{author}{\bibinfo{person}{Xiaoqi Jiao}, \bibinfo{person}{Yichun Yin},
  \bibinfo{person}{Lifeng Shang}, \bibinfo{person}{Xin Jiang},
  \bibinfo{person}{Xiao Chen}, \bibinfo{person}{Linlin Li},
  \bibinfo{person}{Fang Wang}, {and} \bibinfo{person}{Qun Liu}.}
  \bibinfo{year}{2020}\natexlab{}.
\newblock \showarticletitle{{T}iny{BERT}: Distilling {BERT} for Natural
  Language Understanding}. In \bibinfo{booktitle}{\emph{Findings of the
  Association for Computational Linguistics: EMNLP 2020}}.
  \bibinfo{publisher}{Association for Computational Linguistics},
  \bibinfo{address}{Online}, \bibinfo{pages}{4163--4174}.
\newblock
\urldef\tempurl%
\url{https://doi.org/10.18653/v1/2020.findings-emnlp.372}
\showDOI{\tempurl}


\bibitem[\protect\citeauthoryear{Khattab and Zaharia}{Khattab and
  Zaharia}{2020}]%
        {khattab2020colbert}
\bibfield{author}{\bibinfo{person}{Omar Khattab} {and} \bibinfo{person}{Matei
  Zaharia}.} \bibinfo{year}{2020}\natexlab{}.
\newblock \showarticletitle{Colbert: Efficient and effective passage search via
  contextualized late interaction over bert}. In
  \bibinfo{booktitle}{\emph{Proceedings of the 43rd International ACM SIGIR
  conference on research and development in Information Retrieval}}.
  \bibinfo{pages}{39--48}.
\newblock


\bibitem[\protect\citeauthoryear{Kingma and Ba}{Kingma and Ba}{2014}]%
        {kingma2014adam}
\bibfield{author}{\bibinfo{person}{Diederik~P Kingma} {and}
  \bibinfo{person}{Jimmy Ba}.} \bibinfo{year}{2014}\natexlab{}.
\newblock \showarticletitle{Adam: A method for stochastic optimization}.
\newblock \bibinfo{journal}{\emph{arXiv preprint arXiv:1412.6980}}
  (\bibinfo{year}{2014}).
\newblock


\bibitem[\protect\citeauthoryear{Lin, Nogueira, and Yates}{Lin
  et~al\mbox{.}}{2021}]%
        {lin2021pretrained}
\bibfield{author}{\bibinfo{person}{Jimmy Lin}, \bibinfo{person}{Rodrigo
  Nogueira}, {and} \bibinfo{person}{Andrew Yates}.}
  \bibinfo{year}{2021}\natexlab{}.
\newblock \showarticletitle{Pretrained transformers for text ranking: Bert and
  beyond}.
\newblock \bibinfo{journal}{\emph{Synthesis Lectures on Human Language
  Technologies}} \bibinfo{volume}{14}, \bibinfo{number}{4}
  (\bibinfo{year}{2021}), \bibinfo{pages}{1--325}.
\newblock


\bibitem[\protect\citeauthoryear{Llordes, Ganguly, Bhatia, and Agarwal}{Llordes
  et~al\mbox{.}}{2023}]%
        {llordes2023explain}
\bibfield{author}{\bibinfo{person}{Michael Llordes}, \bibinfo{person}{Debasis
  Ganguly}, \bibinfo{person}{Sumit Bhatia}, {and} \bibinfo{person}{Chirag
  Agarwal}.} \bibinfo{year}{2023}\natexlab{}.
\newblock \showarticletitle{Explain like I am BM25: Interpreting a Dense
  Model's Ranked-List with a Sparse Approximation}.
\newblock \bibinfo{journal}{\emph{arXiv preprint arXiv:2304.12631}}
  (\bibinfo{year}{2023}).
\newblock


\bibitem[\protect\citeauthoryear{MacAvaney, Nardini, Perego, Tonellotto,
  Goharian, and Frieder}{MacAvaney et~al\mbox{.}}{2020}]%
        {macavaney2020expansion}
\bibfield{author}{\bibinfo{person}{Sean MacAvaney},
  \bibinfo{person}{Franco~Maria Nardini}, \bibinfo{person}{Raffaele Perego},
  \bibinfo{person}{Nicola Tonellotto}, \bibinfo{person}{Nazli Goharian}, {and}
  \bibinfo{person}{Ophir Frieder}.} \bibinfo{year}{2020}\natexlab{}.
\newblock \showarticletitle{Expansion via prediction of importance with
  contextualization}. In \bibinfo{booktitle}{\emph{Proceedings of the 43rd
  International ACM SIGIR conference on research and development in Information
  Retrieval}}. \bibinfo{pages}{1573--1576}.
\newblock


\bibitem[\protect\citeauthoryear{Maia, Handschuh, Freitas, Davis, McDermott,
  Zarrouk, and Balahur}{Maia et~al\mbox{.}}{2018}]%
        {maia201818}
\bibfield{author}{\bibinfo{person}{Macedo Maia}, \bibinfo{person}{Siegfried
  Handschuh}, \bibinfo{person}{Andr{\'e} Freitas}, \bibinfo{person}{Brian
  Davis}, \bibinfo{person}{Ross McDermott}, \bibinfo{person}{Manel Zarrouk},
  {and} \bibinfo{person}{Alexandra Balahur}.} \bibinfo{year}{2018}\natexlab{}.
\newblock \showarticletitle{Www'18 open challenge: financial opinion mining and
  question answering}. In \bibinfo{booktitle}{\emph{Companion proceedings of
  the the web conference 2018}}. \bibinfo{pages}{1941--1942}.
\newblock


\bibitem[\protect\citeauthoryear{Nguyen, Rosenberg, Song, Gao, Tiwary,
  Majumder, and Deng}{Nguyen et~al\mbox{.}}{2016}]%
        {nguyen2016ms}
\bibfield{author}{\bibinfo{person}{Tri Nguyen}, \bibinfo{person}{Mir
  Rosenberg}, \bibinfo{person}{Xia Song}, \bibinfo{person}{Jianfeng Gao},
  \bibinfo{person}{Saurabh Tiwary}, \bibinfo{person}{Rangan Majumder}, {and}
  \bibinfo{person}{Li Deng}.} \bibinfo{year}{2016}\natexlab{}.
\newblock \showarticletitle{MS MARCO: A human generated machine reading
  comprehension dataset}. In \bibinfo{booktitle}{\emph{CoCo@ NIPs}}.
\newblock


\bibitem[\protect\citeauthoryear{Nogueira, Jiang, and Lin}{Nogueira
  et~al\mbox{.}}{2020}]%
        {nogueira2020document}
\bibfield{author}{\bibinfo{person}{Rodrigo Nogueira}, \bibinfo{person}{Zhiying
  Jiang}, {and} \bibinfo{person}{Jimmy Lin}.} \bibinfo{year}{2020}\natexlab{}.
\newblock \showarticletitle{Document ranking with a pretrained
  sequence-to-sequence model}.
\newblock \bibinfo{journal}{\emph{arXiv preprint arXiv:2003.06713}}
  (\bibinfo{year}{2020}).
\newblock


\bibitem[\protect\citeauthoryear{Paszke, Gross, Chintala, Chanan, Yang, DeVito,
  Lin, Desmaison, Antiga, and Lerer}{Paszke et~al\mbox{.}}{2017}]%
        {paszke2017automatic}
\bibfield{author}{\bibinfo{person}{Adam Paszke}, \bibinfo{person}{Sam Gross},
  \bibinfo{person}{Soumith Chintala}, \bibinfo{person}{Gregory Chanan},
  \bibinfo{person}{Edward Yang}, \bibinfo{person}{Zachary DeVito},
  \bibinfo{person}{Zeming Lin}, \bibinfo{person}{Alban Desmaison},
  \bibinfo{person}{Luca Antiga}, {and} \bibinfo{person}{Adam Lerer}.}
  \bibinfo{year}{2017}\natexlab{}.
\newblock \showarticletitle{Automatic differentiation in pytorch}.
\newblock  (\bibinfo{year}{2017}).
\newblock


\bibitem[\protect\citeauthoryear{Peng, Ding, Zhong, Shen, Liu, Zhang, Ouyang,
  and Tao}{Peng et~al\mbox{.}}{2023}]%
        {peng2023towards}
\bibfield{author}{\bibinfo{person}{Keqin Peng}, \bibinfo{person}{Liang Ding},
  \bibinfo{person}{Qihuang Zhong}, \bibinfo{person}{Li Shen},
  \bibinfo{person}{Xuebo Liu}, \bibinfo{person}{Min Zhang},
  \bibinfo{person}{Yuanxin Ouyang}, {and} \bibinfo{person}{Dacheng Tao}.}
  \bibinfo{year}{2023}\natexlab{}.
\newblock \showarticletitle{Towards Making the Most of ChatGPT for Machine
  Translation}.
\newblock \bibinfo{journal}{\emph{arXiv preprint arXiv:2303.13780}}
  (\bibinfo{year}{2023}).
\newblock


\bibitem[\protect\citeauthoryear{Robertson and Walker}{Robertson and
  Walker}{1994}]%
        {robertson1994some}
\bibfield{author}{\bibinfo{person}{Stephen~E Robertson} {and}
  \bibinfo{person}{Steve Walker}.} \bibinfo{year}{1994}\natexlab{}.
\newblock \showarticletitle{Some simple effective approximations to the
  2-poisson model for probabilistic weighted retrieval}. In
  \bibinfo{booktitle}{\emph{SIGIR’94}}. Springer, \bibinfo{pages}{232--241}.
\newblock


\bibitem[\protect\citeauthoryear{Sallam, Salim, Barakat, and Al-Tammemi}{Sallam
  et~al\mbox{.}}{2023}]%
        {sallam2023chatgpt}
\bibfield{author}{\bibinfo{person}{Malik Sallam}, \bibinfo{person}{Nesreen
  Salim}, \bibinfo{person}{Muna Barakat}, {and} \bibinfo{person}{Alaa
  Al-Tammemi}.} \bibinfo{year}{2023}\natexlab{}.
\newblock \showarticletitle{ChatGPT applications in medical, dental, pharmacy,
  and public health education: A descriptive study highlighting the advantages
  and limitations}.
\newblock \bibinfo{journal}{\emph{Narra J}} \bibinfo{volume}{3},
  \bibinfo{number}{1} (\bibinfo{year}{2023}), \bibinfo{pages}{e103--e103}.
\newblock


\bibitem[\protect\citeauthoryear{Sanderson and Croft}{Sanderson and
  Croft}{2012}]%
        {sanderson2012history}
\bibfield{author}{\bibinfo{person}{Mark Sanderson} {and}
  \bibinfo{person}{W~Bruce Croft}.} \bibinfo{year}{2012}\natexlab{}.
\newblock \showarticletitle{The history of information retrieval research}.
\newblock \bibinfo{journal}{\emph{Proc. IEEE}} \bibinfo{volume}{100},
  \bibinfo{number}{Special Centennial Issue} (\bibinfo{year}{2012}),
  \bibinfo{pages}{1444--1451}.
\newblock


\bibitem[\protect\citeauthoryear{Sun, Yan, Ma, Ren, Yin, and Ren}{Sun
  et~al\mbox{.}}{2023}]%
        {sun2023chatgpt}
\bibfield{author}{\bibinfo{person}{Weiwei Sun}, \bibinfo{person}{Lingyong Yan},
  \bibinfo{person}{Xinyu Ma}, \bibinfo{person}{Pengjie Ren},
  \bibinfo{person}{Dawei Yin}, {and} \bibinfo{person}{Zhaochun Ren}.}
  \bibinfo{year}{2023}\natexlab{}.
\newblock \showarticletitle{Is ChatGPT Good at Search? Investigating Large
  Language Models as Re-Ranking Agent}.
\newblock \bibinfo{journal}{\emph{arXiv preprint arXiv:2304.09542}}
  (\bibinfo{year}{2023}).
\newblock


\bibitem[\protect\citeauthoryear{Sun}{Sun}{2023}]%
        {sun2023short}
\bibfield{author}{\bibinfo{person}{Zhongxiang Sun}.}
  \bibinfo{year}{2023}\natexlab{}.
\newblock \showarticletitle{A short survey of viewing large language models in
  legal aspect}.
\newblock \bibinfo{journal}{\emph{arXiv preprint arXiv:2303.09136}}
  (\bibinfo{year}{2023}).
\newblock


\bibitem[\protect\citeauthoryear{Susnjak}{Susnjak}{2023}]%
        {susnjak2023applying}
\bibfield{author}{\bibinfo{person}{Teo Susnjak}.}
  \bibinfo{year}{2023}\natexlab{}.
\newblock \showarticletitle{Applying BERT and ChatGPT for Sentiment Analysis of
  Lyme Disease in Scientific Literature}.
\newblock \bibinfo{journal}{\emph{arXiv preprint arXiv:2302.06474}}
  (\bibinfo{year}{2023}).
\newblock


\bibitem[\protect\citeauthoryear{Wang and Komatsuzaki}{Wang and
  Komatsuzaki}{2021}]%
        {gpt-j}
\bibfield{author}{\bibinfo{person}{Ben Wang} {and} \bibinfo{person}{Aran
  Komatsuzaki}.} \bibinfo{year}{2021}\natexlab{}.
\newblock \bibinfo{title}{{GPT-J-6B: A 6 Billion Parameter Autoregressive
  Language Model}}.
\newblock
  \bibinfo{howpublished}{\url{https://github.com/kingoflolz/mesh-transformer-jax}}.
\newblock


\bibitem[\protect\citeauthoryear{Wang, Yang, and Wei}{Wang
  et~al\mbox{.}}{2023b}]%
        {wang2023query2doc}
\bibfield{author}{\bibinfo{person}{Liang Wang}, \bibinfo{person}{Nan Yang},
  {and} \bibinfo{person}{Furu Wei}.} \bibinfo{year}{2023}\natexlab{b}.
\newblock \bibinfo{title}{Query2doc: Query Expansion with Large Language
  Models}.
\newblock
\newblock
\showeprint[arxiv]{2303.07678}~[cs.IR]


\bibitem[\protect\citeauthoryear{Wang, Wei, Dong, Bao, Yang, and Zhou}{Wang
  et~al\mbox{.}}{2020}]%
        {wang2020minilm}
\bibfield{author}{\bibinfo{person}{Wenhui Wang}, \bibinfo{person}{Furu Wei},
  \bibinfo{person}{Li Dong}, \bibinfo{person}{Hangbo Bao}, \bibinfo{person}{Nan
  Yang}, {and} \bibinfo{person}{Ming Zhou}.} \bibinfo{year}{2020}\natexlab{}.
\newblock \showarticletitle{Minilm: Deep self-attention distillation for
  task-agnostic compression of pre-trained transformers}.
\newblock \bibinfo{journal}{\emph{Advances in Neural Information Processing
  Systems}}  \bibinfo{volume}{33} (\bibinfo{year}{2020}),
  \bibinfo{pages}{5776--5788}.
\newblock


\bibitem[\protect\citeauthoryear{Wang, Xie, Ding, Feng, and Xia}{Wang
  et~al\mbox{.}}{2023a}]%
        {wang2023chatgpt}
\bibfield{author}{\bibinfo{person}{Zengzhi Wang}, \bibinfo{person}{Qiming Xie},
  \bibinfo{person}{Zixiang Ding}, \bibinfo{person}{Yi Feng}, {and}
  \bibinfo{person}{Rui Xia}.} \bibinfo{year}{2023}\natexlab{a}.
\newblock \showarticletitle{Is ChatGPT a Good Sentiment Analyzer? A Preliminary
  Study}.
\newblock \bibinfo{journal}{\emph{arXiv preprint arXiv:2304.04339}}
  (\bibinfo{year}{2023}).
\newblock


\bibitem[\protect\citeauthoryear{Wolf, Debut, Sanh, Chaumond, Delangue, Moi,
  Cistac, Rault, Louf, Funtowicz, et~al\mbox{.}}{Wolf et~al\mbox{.}}{2019}]%
        {wolf2019huggingface}
\bibfield{author}{\bibinfo{person}{Thomas Wolf}, \bibinfo{person}{Lysandre
  Debut}, \bibinfo{person}{Victor Sanh}, \bibinfo{person}{Julien Chaumond},
  \bibinfo{person}{Clement Delangue}, \bibinfo{person}{Anthony Moi},
  \bibinfo{person}{Pierric Cistac}, \bibinfo{person}{Tim Rault},
  \bibinfo{person}{R{\'e}mi Louf}, \bibinfo{person}{Morgan Funtowicz},
  {et~al\mbox{.}}} \bibinfo{year}{2019}\natexlab{}.
\newblock \showarticletitle{Huggingface's transformers: State-of-the-art
  natural language processing}.
\newblock \bibinfo{journal}{\emph{arXiv preprint arXiv:1910.03771}}
  (\bibinfo{year}{2019}).
\newblock


\bibitem[\protect\citeauthoryear{Yang, Yih, and Meek}{Yang
  et~al\mbox{.}}{2015}]%
        {yang2015wikiqa}
\bibfield{author}{\bibinfo{person}{Yi Yang}, \bibinfo{person}{Wen-tau Yih},
  {and} \bibinfo{person}{Christopher Meek}.} \bibinfo{year}{2015}\natexlab{}.
\newblock \showarticletitle{Wikiqa: A challenge dataset for open-domain
  question answering}. In \bibinfo{booktitle}{\emph{Proceedings of the 2015
  conference on empirical methods in natural language processing}}.
  \bibinfo{pages}{2013--2018}.
\newblock


\bibitem[\protect\citeauthoryear{Zhang, Ding, and Jing}{Zhang
  et~al\mbox{.}}{2022}]%
        {zhang2022would}
\bibfield{author}{\bibinfo{person}{Bowen Zhang}, \bibinfo{person}{Daijun Ding},
  {and} \bibinfo{person}{Liwen Jing}.} \bibinfo{year}{2022}\natexlab{}.
\newblock \showarticletitle{How would Stance Detection Techniques Evolve after
  the Launch of ChatGPT?}
\newblock \bibinfo{journal}{\emph{arXiv preprint arXiv:2212.14548}}
  (\bibinfo{year}{2022}).
\newblock


\bibitem[\protect\citeauthoryear{Zhang, Liu, and Zhang}{Zhang
  et~al\mbox{.}}{2023}]%
        {zhang2023extractive}
\bibfield{author}{\bibinfo{person}{Haopeng Zhang}, \bibinfo{person}{Xiao Liu},
  {and} \bibinfo{person}{Jiawei Zhang}.} \bibinfo{year}{2023}\natexlab{}.
\newblock \showarticletitle{Extractive Summarization via ChatGPT for Faithful
  Summary Generation}.
\newblock \bibinfo{journal}{\emph{arXiv preprint arXiv:2304.04193}}
  (\bibinfo{year}{2023}).
\newblock


\bibitem[\protect\citeauthoryear{Zhang and Sabuncu}{Zhang and Sabuncu}{2018}]%
        {zhang2018generalized}
\bibfield{author}{\bibinfo{person}{Zhilu Zhang} {and} \bibinfo{person}{Mert
  Sabuncu}.} \bibinfo{year}{2018}\natexlab{}.
\newblock \showarticletitle{Generalized cross entropy loss for training deep
  neural networks with noisy labels}.
\newblock \bibinfo{journal}{\emph{Advances in neural information processing
  systems}}  \bibinfo{volume}{31} (\bibinfo{year}{2018}).
\newblock


\bibitem[\protect\citeauthoryear{Zhuang, Li, and Zuccon}{Zhuang
  et~al\mbox{.}}{2021}]%
        {zhuang2021deep}
\bibfield{author}{\bibinfo{person}{Shengyao Zhuang}, \bibinfo{person}{Hang Li},
  {and} \bibinfo{person}{Guido Zuccon}.} \bibinfo{year}{2021}\natexlab{}.
\newblock \showarticletitle{Deep query likelihood model for information
  retrieval}. In \bibinfo{booktitle}{\emph{European Conference on Information
  Retrieval}}. Springer, \bibinfo{pages}{463--470}.
\newblock


\bibitem[\protect\citeauthoryear{Zhuang and Zuccon}{Zhuang and Zuccon}{2021}]%
        {zhuang2021tilde}
\bibfield{author}{\bibinfo{person}{Shengyao Zhuang} {and}
  \bibinfo{person}{Guido Zuccon}.} \bibinfo{year}{2021}\natexlab{}.
\newblock \showarticletitle{TILDE: Term independent likelihood moDEl for
  passage re-ranking}. In \bibinfo{booktitle}{\emph{Proceedings of the 44th
  International ACM SIGIR Conference on Research and Development in Information
  Retrieval}}. \bibinfo{pages}{1483--1492}.
\newblock


\end{thebibliography}
\end{document}